\documentclass[twocolumn]{emulateapj}

\usepackage{apjfonts}

\def\ltsima{$\; \buildrel < \over \sim \;$}
\def\gtsima{$\; \buildrel > \over \sim \;$}
\def\lsim{\lower.5ex\hbox{\ltsima}}
\def\gsim{\lower.5ex\hbox{\gtsima}}

\slugcomment{Accepted for publication in The Astrophysical Journal Letters}

\shorttitle{Blue stragglers in Pal14} 
\shortauthors{Beccari et al.}

\begin{document}


\title{The non-segregated population of blue straggler stars in the
  remote globular cluster Palomar 14}


\author{Giacomo Beccari\altaffilmark{1}, 
Antonio Sollima\altaffilmark{2}, 
Francesco R. Ferraro\altaffilmark{3},
Barbara Lanzoni\altaffilmark{3}, 
Michele Bellazzini\altaffilmark{4}, 
Guido De Marchi\altaffilmark{5}, 
David Valls-Gabaud\altaffilmark{6} \and  
Robert T. Rood\altaffilmark{7}
}

\altaffiltext{1}{European Southern Observatory,
  Karl-Schwarzschild-Strasse 2, 85748 Garching bei M\"unchen, Germany,
  e-mail:gbeccari@eso.org}
\altaffiltext{2}{INAF - Osservatorio Astronomico di Padova, Vicolo
  dell'Osservatorio 5, 35122 Padova, Italy}
\altaffiltext{3}{Dipartimento di Astronomia, Universit\`a degli Studi
  di Bologna, via Ranzani 1, I-40127 Bologna, Italy}
\altaffiltext{4}{INAF - Osservatorio Astronomico di Bologna, Via
  Ranzani 1, 40127 Bologna, Italy}
\altaffiltext{5}{ESA, Space Science Department, Keplerlaan 1, 2200 AG
  Noordwijk, The Netherlands}
\altaffiltext{6}{LERMA, CNRS UMR 8112, Observatoire de Paris, 61 Avenue de 
l'Observatoire, 75014 Paris, France}
\altaffiltext{7}{Astronomy Department, University of Virginia,
 P.O. Box 400325, Charlottesville, VA, 22904,USA}

%%%%%%%%%%%%%%%%%%%%% ABSTRACT

\begin{abstract}

We used deep wide-field observations obtained with the
Canada-France-Hawaii Telescope to study the blue straggler star (BSS)
population in the innermost five arcminutes of the remote Galactic
globular cluster Palomar 14. The BSS radial distribution is found to
be consistent with that of the normal cluster stars, showing no
evidence of central segregation.  Palomar 14 is the third system in
the Galaxy (in addition to $\omega$ Centauri and NGC 2419) showing a
population of BSS not centrally segregated. This is the most direct
evidence that in Palomar 14 two-body relaxation has not fully
established energy equipartition yet, even in the central regions (in
agreement with the estimated half-mass relaxation time, which is
significantly larger than the cluster age). These observational facts
have important implications for the interpretation of the shape of the
mass function and the existence of the tidal tails recently discovered
in this cluster.

\end{abstract}

%%%%%%%%%%%%%%%%%%%%% KEYWORDS


%%%%%%%%%%%%%%%%%%%%% INTRODUCTION

\section{Introduction}
\label{sec_intro}

Blue straggler stars (BSS) are hydrogen-burning stars located bluer
and brighter than the main sequence (MS) turn-off (TO) point in the
optical color-magnitude diagram (CMD) of star clusters.  There is
a general consensus in considering BSS as the most massive (with
$M_{\rm BSS}=1-1.4M_{\odot}$; see Shara et al. 1997, Ferraro et al
2006a) luminous objects in the CMD of a globular cluster (GC).
Two physical mechanisms (both affecting and affected by the dynamical
processes occurring in the cluster) have been proposed for their
formation: direct stellar collisions and mass transfer activity in
binary systems \citep{colbss, mtbss1}.  The former is expected to be
particularly important in high-density environments, while it should
be less efficient in loose GCs and in the cluster external regions,
with respect to the undisturbed evolution of primordial binaries.  On
the other hand, the discovery \citep{fe09} of two distinct and
parallel BSS sequences in M30, possibly populated by objects generated
by the two formation processes during the core collapse phase, further
supports the idea that BSS of different origins can coexist within the
same stellar system.  Observational proof of the connection
between the binary and the BSS populations in low-density environments
is testified by the correlation between the number fraction of these
two species recently found in the core of 13 loose GCs \citep{so08}.
In addition, being more massive than the average, BSS tend to sink
toward the bottom of the cluster potential well, under the action of 
energy equipartition. In most of the surveyed GCs ($\sim20$ to date)
BSS appear to be strongly concentrated in the core
(see, e.g., Figure 2 in Ferraro et al. 2003 and Figure 6 in
Ferraro \& Lanzoni 2009), with their fraction, measured with respect to 
ordinary cluster stars, decreasing at increasing distance from the center
and showing an upturn in the external regions \citep[see also][]{fe93, fe04,dal09}.  
This feature has been interpreted by means of numerical
simulations~\citep[][]{ma04,ma06,la07a,la07b} 
showing that the central peak is due both to BSS formed in place
because of stellar collisions, and to mass transfer BSS sunk to the
centre under the effect of mass segregation.  In contrast, the rising
branch in the cluster outskirts is due to BSS generated by the
unperturbed evolution of primordial binaries, which are preferentially
orbiting in regions where the dynamical friction timescale is longer
than the cluster age. The only two exceptions currently known are
 $\omega$Centauri \citep{fe06} and NGC 2419 \citep{da08}, where BSS show the
same radial distribution as that of the other cluster stars. 
This fact indicates that these two GCs have
not yet reached a status of energy equipartition, and their BSS result
from the evolution of binary systems whose radial distribution has not
been altered by the the process of mass segregation.

Based on these results, here we use the BSS radial distribution to
investigate the dynamical state of the remote GC Palomar 14 (hereafter
Pal14), located in the outer Halo of the Milky Way, at a distance of
$\sim71$ kpc \citep[][hereafter S11]{so11}.  

Pal14 has been indicated as
one of the best candidates to test alternative theories of gravity
\citep{bau05, solnip10} and recent analyses have shown a peculiar
structure and kinematics of this cluster. Adopting a sample of 
17 stars \citet{jo09} measured a very small
velocity dispersion ($\sigma_{v}=0.38\pm0.12$ km/s). \citet{ku10}
claimed that such a small value is not compatible with the presence of
the fraction of binaries ($>20\%$) expected in a loose
GC. They argued that either this cluster hosts a low
fraction of binaries, or it constitutes a ``deep-freeze'' with an
unusually low velocity dispersion.  By measuring the cluster mass
function between 0.53 and 0.78$M_\odot$, \citet{jo09} found a
significantly flatter slope than the canonical value and concluded
that either Pal14 formed with only few low-mass stars, or it is mass
segregated and lost most of its low-mass stars through interaction
with the Galactic tidal field. Indeed, two well defined and extended
tidal tails associated with Pal14, likely due to an active process of
tidal stripping, have been recently detected (S11). However, no
measurement of the degree of mass segregation was available to date
and the estimated half-mass relaxation time ($t_{rh}\sim 20$ Gyr; S11)
is much longer than the cluster age \citep[$t\sim 10.5$
Gyr;][]{dot08}. On the basis of these results and using 
N-body simulations, \citet{zonoo11} concluded that either a primordial
mass segregation or a non-canonical initial mass function must have
been established in this cluster after the initial gas expulsion.

With the aim of clarifying the dynamical state of Pal14, here we study
the radial distribution of its BSS population. In Sect. 2 the adopted
photometric dataset is presented. In Section 3 and 4 the
BSS and the reference populations are defined and their radial
distribution is derived and analyzed. In Section 5 we present our
conclusions.

%%%%%%%%%%%%%%%%%%%%% SECTION

\section{Catalogue and completeness level}
\label{sec_cata}
\begin{figure}[t] 
\centering 
\includegraphics[scale=0.44]{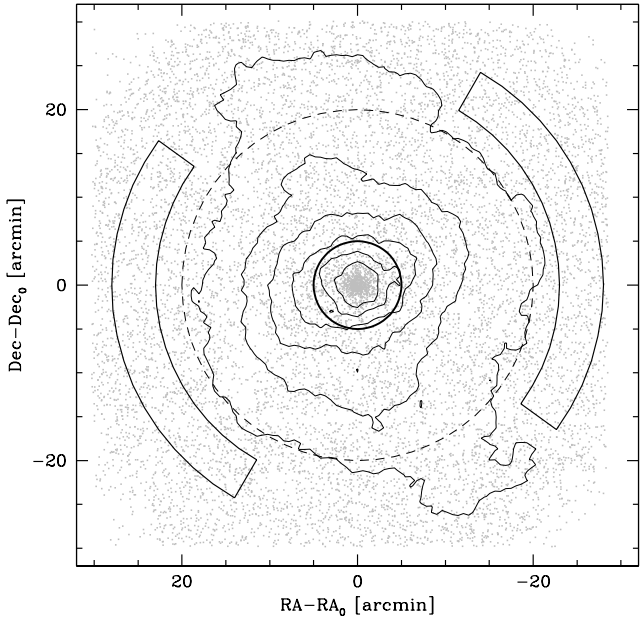}
\caption{Map of the MegaCam dataset of Palomar 14. The region
  considered in the present analysis ($r<300\arcsec\simeq 2\, r_h
  \simeq 8\, r_c$) is marked with the thick solid circle. The two
  regions contained within the thick solid lines have been used to
  define the sample of Galactic field stars. The tidal radius
  ($r_t=20\arcmin$) is shown for comparison and marked with the dashed
  circle.  The iso-density contours of the large-scale surface density
  around Pal14, showing the tidal tails discovered by S11, are plotted
  with thin solid lines.  }
\label{map}
\end{figure}

This work is based on observations performed with the wide-field
camera MegaCam, mounted at the Canada-France-Hawaii Telescope (CFHT).
The photometric dataset and the adopted reduction procedures are
described in S11.  The final catalogue includes 40,000
objects sampled in the $g$' and $r$' filters over an area of 1 deg$^2$
centered on the cluster.  A map of the MegaCam
dataset is shown in Fig. \ref{map} and the CMD for the innermost
$300\arcsec$ of the cluster is shown in the left panel of Fig. \ref{cmd}.

The loose structure of Pal14 guarantees that the photometric
analysis is accurate and complete even in the very central regions of
the cluster.  However, in order to quantitatively estimate the level
of completeness of the photometric sample in the central region of the
cluster, we performed a detailed comparison with
high-resolution images of the cluster obtained with the Hubble Space
Telescope (HST). We retrieved a set of deep, multi-band
images (with total integration times of 8540s and of 10320s in the
$F555W$ and $F814W$ filers, respectively) secured with the Wide Field
Planetary Camera 2 (WFPC2, GO-6512; PI: Hesser).  We analyzed these
images by using DAOPHOTII \citep{ste87}. Briefly, an accurate Point
Spread Function (PSF) was estimated on each frame and adopted for the
first run of the PSF fitting procedure.  A master list of stars was
extracted from a deep, high signal-to-noise image obtained from the
montage of the entire dataset.  
Then, the average of the magnitudes
measured \citep[through ALLFRAME;][]{ste87} in each single frame for
every master list object was adopted as the star magnitude in the
final catalogue, and the error of the mean was assumed to be the
associated photometric error.  
The WFPC2 catalogue includes 2766 stars with $\sigma_{mag}<0.2$
and sharpness parameter $|sh|<0.2$ down to $F555W\sim28$, i.e. 5 mag 
below the MS-TO.  The number of sampled stars is
fully compatible with the 2752 stars sampled by~\citet{jo09} and
the completeness study shown in their Fig. 4, safely assumed as representative
of the quality of our WFPC2 photometry, indicates that our WFPC2 catalogue is 
$100\%$ complete well below the MS-TO.    
The completeness of the MegaCam catalogue is quantified as the
fraction of stars in common with the WFPC2 catalogue in a given
 magnitude interval, with respect to the total number of stars
sampled by the WFPC2 in the same range. 
We find $93\%$, $95\%$ and $100\%$ of stars at $F555W\sim g'\leq23.5, 23$ and 22.5,
respectively.

%%%%%%%%%%%%%%%%%%%%%%SUBSECTION
\subsection{Center of Gravity}

By following the procedure described in \citet[][see also Dalessandro
  et al. 2008]{mo95}, we have estimated the center of gravity
($C_{\rm grav}$) of Pal14 as the barycenter of the resolved stars.  To
this aim, we first performed a rough selection of the cluster stars
along the canonical evolutionary sequences in the
CMD. Then, through a
sigma clipping procedure, we averaged the $\alpha$ and $\delta$
positions of all the stars contained within three circular areas of
radius $r<60\arcsec,70\arcsec$ and $80\arcsec$ around the centre
quoted by \citet{h96}. Three barycenters were measured in each area by
using stars brighter than $g$'$= 23.5, 23.7, 24$.  The mean of these
nine measures of the barycenters turns out to be
$\alpha_{J2000}=16^{\rm h} 11^{\rm m} 0.8^{\rm s}$,
$\delta_{J2000}=14^\circ 57\arcmin 27.9\arcsec$, with standard
deviations $3.5\arcsec$ and $1.3\arcsec$ in RA and Dec, respectively
(a large scatter is expected because of the extremely low stellar
density even in the core region of the cluster). This position of 
the cluster center is in agreement (within the errors 
$\Delta\alpha\simeq3.19\arcsec$, $\Delta\delta\simeq -0.05$) 
with the one found by~\citet{hilk06}.

%%%%%%%%%%%%%%%%%%%%% SECTION

\section{Population selection and BSS radial distribution}
\label{sec_pop}

In order to investigate the cluster dynamical state, we studied the BSS
radial distribution and compared it to that of red giant branch (RGB) and horizontal
branch (HB) stars, taken as representative of the ``normal" cluster
population \citep[see][for a recent review]{felan09}.  This requires a
proper selection of the samples in radial annuli around the cluster
centre and an accurate analysis of the contribution due to field
stars located in the foreground/background of the cluster.  Given the
tidal distortion recently detected in Pal14 (S11), its stellar
distribution can be considered to have a spherical symmetry
only for $r\lsim 300\arcsec$ (see Fig. \ref{map}, and Fig. 4 in
S11). This corresponds to $\sim 2 r_h$ (or $\sim 8 r_c$), a distance
where any signature of mass segregation in the radial distribution of
the cluster populations is expected to be well visible. Moreover, at
these radial distances the field contamination is still acceptable
with respect to the number of cluster stars.  Hence, we limit the
following analysis to this portion ($r\le300\arcsec$) of the cluster.
\begin{figure}
 \centering 
 \includegraphics[scale=0.44]{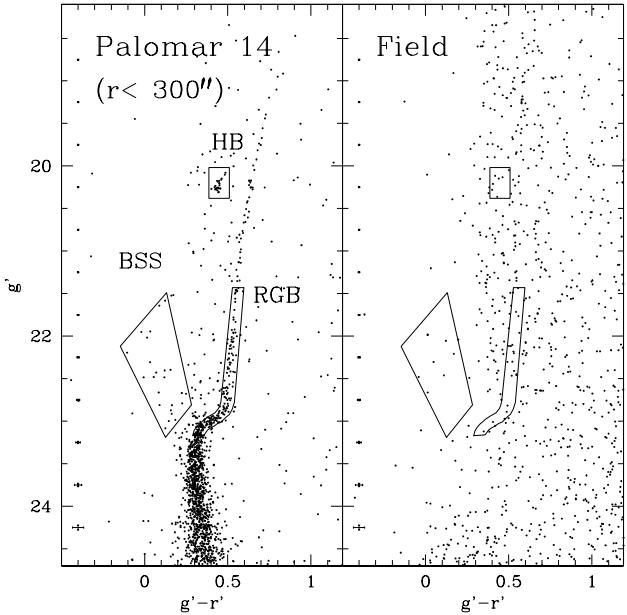}
\caption{{\it Left panel}: CMD of the inner $300\arcsec$ of Pal14, obtained from the
  MegaCam dataset.  The selection boxes used to define the BSS, RGB
  and HB samples are marked with solid lines. The average photometric 
  magnitudes and color errors are shown by the error bars on the left.{\it Right panel}: 
  CMD of the field region ($\sim445$ arcmin$^2$; see Fig.1) 
  with the population selection boxes superimposed.}
\label{cmd}
\end{figure}

In order to define the BSS sample we have considered
the selection box shown in Fig. \ref{cmd}. While \cite{san05} also 
included stars much closer to the MS-TO (see his Fig. 1),
we chose a more conservative limit in order to minimize the risk of contamination by spurious
(blended) sources, which are expected to be very few or zero, but could be critical because of
the small number of stars in Pal14. The same magnitude range was used to
define the RGB sample, which include some sub-giant stars. 
The selection boxes used for the RGB and HB populations 
are shown in Fig. \ref{cmd}.  We
count a total of 24 BSS, 191 RGB and 24 HB
stars.\footnote{\footnotesize{We have carefully checked that slightly 
different assumptions about the selection boxes 
(e.g. using the definitions provided by Sandquist 2005) 
do not change the results of the analysis presented in this paper.}}

\begin{figure} 
\centering 
\includegraphics[scale=0.44]{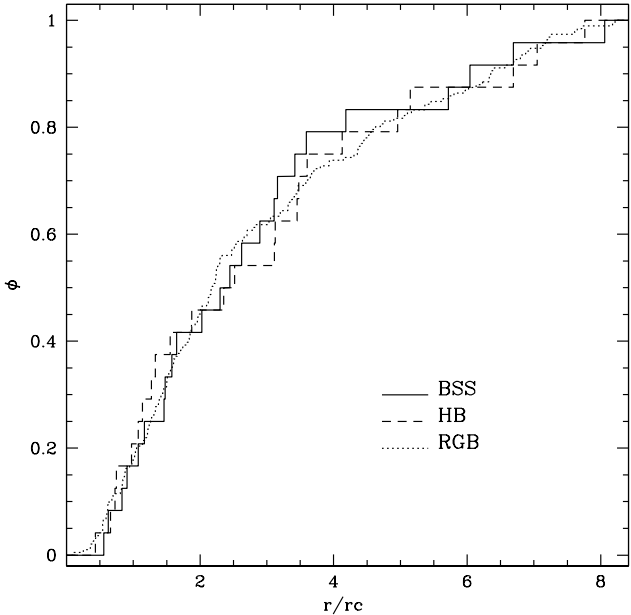}
\caption{Cumulative radial distribution of the BSS, RGB and HB
  populations of Pal14 within $300\arcsec$ from the center, with the
  distance normalized to the cluster core radius ($r_c=36\arcsec$,
  from S11).}
\label{ks}
\end{figure}

The cumulative radial distributions of the three samples are shown in
Fig. \ref{ks}.  No significant difference is found among
the three distributions, thus indicating that none is distinctly
segregated towards the cluster center with respect to the others. A
Kolmogorov-Smirnov test indicates a 99\% probability that the three
samples are extracted from the same parent population.

We also used the Anderson-Darling test, which is more sensitive to the tails of
the empirical cumulative distribution function than Kolmogorov-Smirnov, to assess
possible differences at small and large radial distances. Using the $k$-sample 
variant of the Anderson-Darling test~\citep[][]{sc87}, we found 
that for any pair of samples ($k$=2) the probability that they arise 
from the same underlying distribution is
63\%, while for the combined set ($k$=3) of samples, the probability raises
to 79\%. There is therefore no statistically significant radial segregation between
the samples of BSS, HB and RGB stars.

To further investigate this issue, we studied the radial distribution
of the population ratios by following the procedure described, e.g.,
in \cite{felan09}.  The field of view was divided in three concentric
annuli, each one sampling approximately the same luminosity
fraction. The number of objects belonging to the three populations was
then counted in each annular area (see Table~\ref{tab_count}).  

\begin{table}
\caption{Number of BSS, RGB and HB stars in three concentric annuli
  used in the analysis. The estimated number of contaminating field
  stars for each sample is reported in parenthesis. The mean radius 
  normalized to the cluster core radius is also reported.}
\label{tab_count}    
\centering
\begin{tabular}{c c c c c}       
\hline\hline                 
annulus          & $<r>/r_c $&$N_{\rm BSS}$ & $N_{\rm RGB}$ &  $N_{\rm HB}$\\   
\hline                      
$  0\arcsec- 55\arcsec$ & 0.76  & 8 (0.05) & 64 (0.12) & 9 (0.04)    \\      
$ 55\arcsec-120\arcsec$ & 2.43 & 9 (0.18) & 62 (0.47) & 6 (0.16)    \\
$120\arcsec-300\arcsec$ & 5.83 &7 (1.29) & 65 (3.11) & 9 (1.04)    \\
\hline                                
\end{tabular}
\end{table}

Inorder to estimate the contamination from field stars to the selected
populations we have taken advantage of the wide radial coverage of the
MegaCam catalogue.  Two areas between $23\arcmin$ and $25\arcmin$
(i.e., at $3\arcmin$ from $r_t$) and orthogonal to the direction of
the tidal tail (thus to minimize the possibility that genuine cluster
stars fall in field sample; see Fig. \ref{map}) have been selected as
representative of the Galactic field population. As shown in Fig. \ref{cmd} 
(right panel) no signature of the cluster stellar populations is
found in the field CMD.
We count, respectively, 8, 21 and 7 field stars
within the same BSS, RGB and HB selection boxes discussed above.
We estimate a contamination of $1.8\times10^{-2}$,
$4.7\times10^{-2}$ and $1.6\times10^{-2}$ field stars per square
arcminute, to the BSS, RGB and HB populations, respectively.  By
taking into account the area sampled by the three radial bins, the
number of contaminating field stars in each annulus and for each
population has been derived and it is quoted in parenthesis in Table
\ref{tab_count}. Finally, to compute the average ratio of the 
number of RGB, HB and BSS stars we used Eq. (26) 
in~\citet[][]{ce03} for uncorrelated Poisson variables.

\begin{figure} 
  \centering
   \includegraphics[scale=0.44]{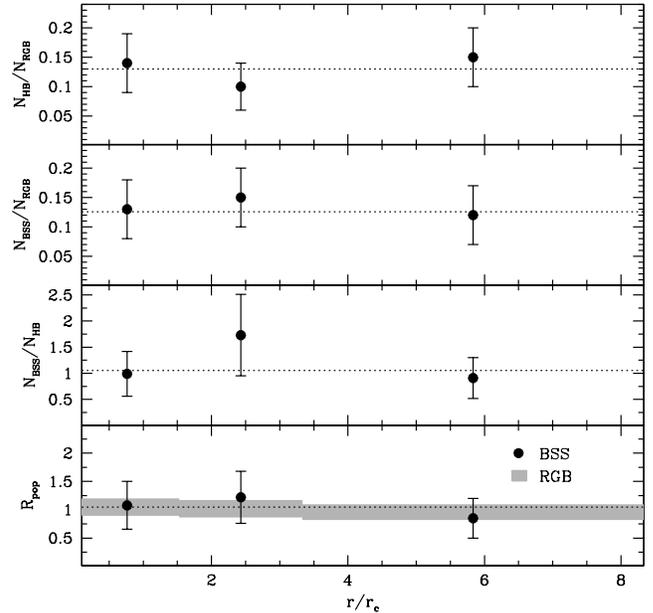}
\caption{From top to bottom: radial distribution of the number
  fractions of HB and RGB stars, BSS and RGB stars, BSS and HB stars,
  and of the double normalized ratios $R_{\rm pop}$ (see text) of BSS
  (dots) and RGB stars (gray regions, with the vertical size
  corresponding to the error bars). The ratios and their errors were 
  calculated assuming that the number counts of each sample
were uncorrelated Poisson variables~\citep[][]{ce03}. }
  \label{radist}
\end{figure}

As shown in Fig. \ref{radist}
(upper panel), the radial distribution of the ratio between HB and RGB
stars is flat, in agreement with what expected for normal cluster
stars, which all have (essentially) the same mass. Consistently with
what found above, and despite their higher mass, also BSS are
distributed as the normal cluster population and show no evidence of
central segregation (Fig. \ref{radist}, second and third panels).

We have also investigated the radial behavior of the {\it
  double normalized ratio} \citep{fe93}, which reports, per each
considered annulus, the fraction of stars observed in a given
evolutionary stage (BSS, HB, RGB) divided by the faction of cluster
light sampled in that annulus by the observations: $R_{\rm
  pop}=(N^{\rm pop}_{\rm samp}/N^{\rm pop}_{\rm tot})/(L_{\rm
  samp}/L_{\rm tot})$, with pop=BSS, HB, RGB.  Since the number of
stars in any post-MS stage scales linearly with the total luminosity
of the stellar population \citep{re88}, this ratio is predicted to be
equal to 1 for any (not segregated) reference populations, while a
larger (smaller) value is expected for populations which are more
(less) concentrated.  In order to estimate the sampled light, we have
integrated the King profile best-fitting the surface density
distribution (from S11).  The radial trend of $R_{\rm RGB}$
and $R_{\rm BSS}$ in the same annuli previously defined is
shown in the lower panel of Fig. \ref{radist}.  As apparent, it
turns out to be constant and equal to 1 not only for the RGB (and the HB)
stars, but also for the BSS, thus further demonstrating that this
population has a radial distribution fully consistent with that of the
reference ones.

\section{Discussion}

The radial distribution of BSS in Pal14 is indistinguishable from that
of its normal (and less massive) stars. As in the case of the only two
other clusters showing the same feature \citep[$\omega$Cen and
 NGC2419;][respectively]{fe06,da08}, this is an observational
proof that Pal14 is dynamically young, still far from having 
established energy equipartition even in its innermost regions. This is in
agreement with the extremely long half-mass radius relaxation time
($\sim 20$ Gyr) recently estimated by S11.  Moreover, our results
suggest that the unusually flat mass function measured by \citet{jo09}
cannot be explained by energy equipartition developed during the
cluster dynamical evolution, but should be primordial \citep[as
suggested by][]{zonoo11}. We note that since the mass range
covered by that study is quite limited, further investigation is
needed.  Finally, the negligible degree of
relaxation of Pal14 suggests that the observed tidal tails should not  be
preferentially populated by low mass stars evaporated from the cluster.

The flat BSS radial distribution also suggests that (as expected in
such a low density environment) stellar collisions played a minor role
in generating these exotica and affecting the binary
population. Hence, as in the case of $\omega$Centauri and NGC 2419,
the BSS we are observing likely derive from the evolution of
primordial binaries and can be used to get a rough estimate of the
fraction of such a population. As a first consideration, we note that
the number of BSS normalized to the sampled luminosity \citep[in units
  of $10^4 L_\odot$, see][]{fe95} is $S4_{\rm BSS}\simeq 2$ in
$\omega$Centauri and $S4_{\rm BSS}\simeq 3$ in NGC 2419. The
same ratio in Pal14 rises to 29 (i.e. a value 10 times larger than
what found in the two clusters with similar BSS radial distribution).
However this value is not that surprising when compared to the
field. In fact, as discussed by \citet{fe06}, the observed BSS
specific frequency $N_{\rm BSS}/N_{\rm HB}=0.1$ in $\omega$Centauri
turned out to be $\sim 40$ times lower that what observed in the field
($N_{\rm BSS}/N_{\rm HB}=4$) by Preston \& Sneden (2000).  The value
found in Pal14 is $N_{\rm BSS}/N_{\rm HB}\sim 1$, in much better
agreement with the above field sample.

Under the hypothesis that all the BSS are originated by primordial
binaries, this possibly suggests that the binary fraction in Pal14
(and in the field) might be much higher than in the other two GCs
\citep[in $\omega$Cen it amounts to $f_{\rm bin}\sim
  13\%$;][]{so07a}. A rough estimate of the binary fraction in
Palomar 14 can be derived from the correlation between 
the measured binary fraction and the cluster integrated magnitude 
found by~\citet{so10} in a sample of 18 open and low-density
globular clusters.  From their Figure 7, and adopting $M_V=-4.95$
(S11) we predict $f_{\rm bin}\sim 30-40\%$ for Palomar 14.  In
addition, the comparison between the number of BSS per unit luminosity
\citep[from][]{fe95} and the fraction of binaries \citep{so07b}
measured in a sample of low-density GCs (in which the collisional
channel of BSS formation is expected to be negligible) shows that
while BSS-poor GCs (with $8<S4_{\rm BSS}<13$) host a small fraction of
binaries ($9<f_{\rm bin}<16\%$), the only cluster (Palomar 12) with a
value of $S4_{\rm BSS}$ similar to Pal14 has a binary fraction of
$\sim 40\%$.  Despite the uncertainty affecting these estimates, such
a result and, even more robustly, the existence of a non-collisional
BSS population, suggest that Pal14 might have a non-negligible
fraction (of the order of $\sim 30-40\%$) of binaries.

Deep and high-quality imaging of Pal14 main sequence are urged
for a direct estimate of the binary fraction and a measurement of the
mass function in a wide range of masses. Additional spectroscopic
campaigns able to more precisely estimate the cluster velocity
dispersion would also be very useful to constrain more tightly the
dynamical state of the cluster.

%%%%%%%%%%%%%%%%%%%%% ACKNOWLEDGMENTS

\acknowledgments 
This research is part of the project {\it COSMIC-LAB}
funded by the {\it European Research Council} (under contract
ERC-2010-AdG-267675).  The financial contribution of {\it Istituto
 Nazionale di Astrofisica} (INAF, under contract PRIN-INAF 2008) and
the {\it Agenzia Spaziale Italiana} (under contract ASI/INAF
I/009/10/0) is also acknowledged.
This research was supported by project ANR POMMME (ANR 09-BLAN-0228).
Based on observations obtained with MegaPrime/MegaCam, a joint project 
of CFHT and CEA/DAPNIA, at the CFHT which is operated by the National 
Research Council (NRC) of Canada, the Institut National des Sciences de 
l'Univers of the Centre National de la Recherche Scientifique (CNRS) of 
France, and the University of Hawaii.
%%%%%%%%%%%%%%%%%%%%% BIBLIOGRAPHY

\clearpage

%%%%%%%%%%%%%%%%%% TABLE

%%%%%%%%%%%%%%%%%%%%%%%%% FIGURE

\end{document}